\documentstyle[11pt,epsf]{article}
\newcommand{\Dslash}{D \!\!\!\! /} 
\newcommand{\kslash}{k \!\!\! /} 
\newcommand{\partialslash}{\partial \!\!\! /} 
\newcommand{\half}{\mbox{\small{$\frac{1}{2}$}}} 
\newcommand{\Nf}{N_{\!f}} 
\newcommand{\leqq}{\mbox{\footnotesize{$\stackrel{<}{\sim}$}}} 
\begin{document}
\title{The QCD $\beta$-function at $O(1/N_{\! f})$}  
\author{J.A. Gracey, \\ Department of Applied Mathematics and Theoretical
Physics, \\ University of Liverpool, \\ P.O. Box 147, \\ Liverpool, \\ 
L69 3BX, \\ United Kingdom.}
\date{} 
\maketitle
\vspace{5cm}
\noindent
{\bf Abstract.} The leading order coefficients of the $\beta$-function of QCD
are computed in a large $N_{\! f}$ expansion. They are in agreement with the
three loop $\overline{\mbox{MS}}$ calculation. The method involves computing 
the anomalous dimension of the operator $(G^a_{\mu\nu})^2$ at the 
$d$-dimensional fixed point in the non-abelian Thirring model to which QCD is 
equivalent in this limit. The effect the $O(1/N_{\!f})$ corrections have on the
location of the infrared stable fixed point for a range of $N_{\!f}$ is also 
examined.  

\vspace{-17cm} 
\hspace{13.5cm} 
{\bf LTH 366} 

\newpage
The strong force of the standard model is described by quantum chromodynamics,
(QCD), which is an $SU(3)$ gauge theory which is asymptotically free. At large
energies the fundamental quarks behave as though they were non-interacting. In
terms of the field theory itself this property is a consequence of the leading
coefficient of the $\beta$-function being negative. The initial one loop 
calculation was carried out in \cite{1}. Higher order corrections have also 
been determined. The remaining scheme independent term, ie the two loop 
contribution, was calculated in \cite{2}. The three loop term was computed in 
the Feynman gauge in \cite{3} using dimensional regularization in the 
$\overline{\mbox{MS}}$ scheme. More recently this result was checked in 
\cite{4} in an arbitrary covariant gauge where the remaining renormalization
group functions, such as the gluon anomalous dimension, were also deduced in
arbitrary gauge, \cite{4}. The three loop quark mass anomalous dimension, which
is gauge independent, is available too, \cite{5,6}. One reason for such precise
information is that, for example, it allows one to obtain a more accurate 
insight into the variation of quantities with energy scale. Fundamental in this 
respect is the $\beta$-function as it always appears in the appropriate
renormalization group equation, (rge). 

These higher order analytic calculations of the rge functions are exceedingly 
tedious to compute, however, due to the huge number of Feynman diagrams that 
arise. For such results to be credible it is important to have independent 
checks on the expressions obtained aside from the obvious one of performing
another complete evaluation which may be a waste of resources. In this letter 
we provide the results of such a procedure for the QCD $\beta$-function. This 
is the large $\Nf$ technique of determining exact all orders results of the rge 
functions of gauge theories at successive orders in powers of $1/\Nf$, where 
$\Nf$ is the number of fundamental fields. The technique was initially 
developed for low dimensional models in a series of impressive papers, 
\cite{7,8,9}. Briefly the method involves computing appropriate critical 
exponents at the $d$-dimensional fixed point of the $\beta$-function as $\Nf$ 
$\rightarrow$ $\infty$. Through the critical rge these $d$-dimensional 
exponents encode all orders information on the coefficients of the 
corresponding rge function. Clearly the values will overlap with the lowest 
known orders, providing the partial check we have indicated. From a technical 
point of view one benefit of this approach is the exploitation of the conformal
symmetry at the fixed point which simplifies the resummation of the minimal set
of Feynman diagrams comprising the relevant Schwinger Dyson equation. The 
calculation of the $O(1/\Nf)$ QCD $\beta$-function here completes the leading 
order analysis as the quark, gluon and ghost dimensions were deduced in 
\cite{10}, in the Landau gauge and agreed with the three loop results of 
\cite{11,4}. Another motivation arises from a comment in \cite{4} in regard to 
future calculations. It is indicated that the four loop QCD $\beta$-function is
attainable. The main obstacle, though, would appear to be correctly generating 
and treating the vast numbers of Feynman diagrams. Therefore the new 
coefficients we will deduce from our results will be important in this respect.

We recall that the $O(1/\Nf)$ computation of the QED $\beta$-function is 
available, \cite{12}. That calculation was carried out by inserting the 
implicit bubble sum of the photon propagator in the $2$ and $3$ point functions
and then deducing the $\overline{\mbox{MS}}$ coefficients of the 
renormalization constants using dimensional regularization. Those results have 
been reproduced in the critical point approach, \cite{13}. One interesting 
aspect of \cite{12} was the search for other fixed points in the strictly four 
dimensional QED $\beta$-function, for a range of values of the coupling. 
Although none were observed it would be a worthwhile exercise to repeat that 
analysis in the non-abelian case especially as at two loops such a point 
exists, \cite{14}, for a range of $\Nf$.  

We recall the fundamental ingredients for treating QCD in large $\Nf$ in our 
approach. The lagrangian is 
\begin{equation} 
L ~=~ i \bar{\psi}^{iI} \Dslash \psi^{iI} ~-~ \frac{(G^a_{\mu \nu})^2}{4e^2} 
\end{equation} 
where $\psi^{iI}$ is the quark field, $A^a_\mu$ is the gluon field, the 
covariant derivative is $D_\mu$ $=$ $\partial_\mu$ $+$ $iT^a A_\mu^a$ with 
$T^a$ the generators of the colour gauge group and $G^a_{\mu\nu}$ $=$ 
$\partial_\mu A^a_\nu$ $-$ $\partial_\nu A^a_\mu$ $+$ $f^{abc}A^b_\mu A^c_\nu/e$
is the field strength with $f^{abc}$ the structure constants. The ranges of the
indices are $1$ $\leq$ $i$ $\leq$ $\Nf$, $1$ $\leq$ $a$ $\leq$ $(N^2_c-1)$ and
$1$ $\leq$ $I$ $\leq$ $N_c$ and the Casimirs are $\mbox{tr}(T^aT^b)$ $=$ 
$T(R)\delta^{ab}$, $T^aT^a$ $=$ $C_2(R)$ and $f^{acd}f^{bcd}$ $=$ $C_2(G) 
\delta^{ab}$. To three loops, in $d$-dimensions, [1-4],  
\begin{eqnarray}
\beta(g) &=& (d-4)g + \left[ \frac{2}{3}T(R)\Nf - \frac{11}{6}C_2(G) \right] 
g^2 \nonumber \\
&+& \left[ \frac{1}{2}C_2(R)T(R)\Nf + \frac{5}{6}C_2(G)T(R)\Nf
- \frac{17}{12}C^2_2(G) \right] g^3 \nonumber \\
&-& \left[ \frac{11}{72} C_2(R)T^2(R)\Nf^2 + \frac{79}{432} C_2(G) T^2(R)
\Nf^2 + \frac{1}{16} C^2_2(R) T(R) \Nf \right. \nonumber \\
&&-~ \left. \frac{205}{288}C_2(R)C_2(G)T(R)\Nf 
- \frac{1415}{864} C^2_2(G)T(R)\Nf + \frac{2857}{1728}C^3_2(G) 
\right] g^4 + O(g^5) 
\end{eqnarray} 
where our coupling $g$ is $g$ $=$ $(e/2\pi)^2$. The presence of the $O(g)$ term
of (2), corresponding to the dimension of the coupling in $d$-dimensions, gives
rise to our non-trivial fixed point, $g_c$. Explicitly  
\begin{eqnarray}
g_c &=& \frac{3\epsilon}{T(R)\Nf} + \frac{1}{4T^2(R)\Nf^2} \left[ \frac{}{} 
33C_2(G)\epsilon - \left( 27C_2(R) + 45C_2(G)\right) \epsilon^2 \right.
\nonumber \\ 
&&+~ \left. \left( \frac{99}{4}C_2(R) + \frac{237}{8} C_2(G) \right) 
\epsilon^3 + O(\epsilon^4) \right] + O \left( \frac{1}{\Nf^3} \right)
\end{eqnarray}
where $d$ $=$ $4$ $-$ $2\epsilon$. In the neighbourhood of this point the 
quark and gluon anomalous dimension were deduced in the Landau gauge as, at 
leading order in $1/\Nf$, respectively, \cite{10}, 
\begin{eqnarray}
\eta &=& \frac{C_2(R)\eta^{\mbox{o}}_1}{T(R)\Nf} \\ 
\eta \, + \, \chi &=& - \, \frac{C_2(G) \eta^{\mbox{o}}_1}{2(\mu-2)T(R)\Nf} 
\end{eqnarray}
where $\eta^{\mbox{o}}$ $=$ $(2\mu-1)(\mu-2)\Gamma(2\mu)/[4\Gamma^2(\mu) 
\Gamma(\mu+1) \Gamma(2-\mu)]$ and $d$ $=$ $2\mu$. Moreover, the asymptotic 
scaling forms of the respective propagators are, as $k^2$ $\rightarrow$ 
$\infty$,  
\begin{equation}
\psi(k) ~\sim~ \frac{A\kslash}{(k^2)^{\mu-\alpha}} ~~,~~
A_{\mu\nu}(k) ~\sim~ \frac{B}{(k^2)^{\mu-\beta}}\left[ \eta_{\mu\nu}
- \frac{k_\mu k_\nu}{k^2} \right] 
\end{equation} 
where $\alpha$ $=$ $\mu$ $+$ $\half\eta$, $\beta$ $=$ $1$ $-$ $\eta$ $-$ $\chi$
and $A$ and $B$ are amplitudes though only the combination $z$ $=$ $A^2B$ 
appears in calculations. Specifically $z$ $=$ 
$\Gamma(\mu+1)\eta^{\mbox{o}}/[2(2\mu-1)(\mu-2)T(R)\Nf]$. 

One feature which simplifies the fixed point analysis both for computing the 
$O(1/\Nf)$ corrections to $\beta(g)$ and (4,5) arises from the universality 
class in which QCD belongs. For instance, it is widely accepted that the $O(N)$
bosonic $\sigma$ model and the $O(N)$ $\phi^4$ model are equivalent, or in the
same universality class, at the $d$-dimensional fixed point analogous to (3). 
In other words critical exponents calculated in either field theory at this 
fixed point are the same. So if one wished to examine the critical behaviour of
the three dimensional $O(3)$ Heisenberg ferromagnet to which both are 
equivalent then either model can be used. Another widely studied equivalence is
the Yukawa interaction and the Gross Neveu model with the same chiral 
properties, \cite{15}. Likewise the Thirring model and QED are equivalent and 
have been the subject of recent interest, \cite{16,17}. For QCD the relevant 
model is the non-abelian Thirring model, (NATM), whose lagrangian is  
\begin{equation} 
L ~=~ i \bar{\psi}^{iI} \partialslash \psi^{iI} ~+~ \bar{\psi}^{iI} \gamma^\mu
T^a_{IJ} \psi^{iJ} A^a_\mu ~-~ \frac{(A^a_\mu)^2}{2\lambda} 
\end{equation} 
where $A^a_\mu$ is an auxiliary field, which if eliminated leads to a $4$-fermi
interaction, and $\lambda$ is the coupling whose dimension is $(d-2)$ when 
compared to the $(d-4)$ of the QCD coupling. It has been demonstrated in 
\cite{18} that it is equivalent to QCD in the large $\Nf$ limit. One feature of
that work was the correct reproduction of the three and four gluon vertices of
non-abelian theories, which are absent in (7), by integrating out quark loops
in the gluon $3$- and $4$-point functions. For our computation the major 
simplification is the fact that by calculating with (7) at $g_c$ the resulting
exponents, which are universal, will be equivalent to those computed in (1). 
Then by decoding the exponent using (3), we can deduce information on the 
perturbative structure of the rge functions. A test of this argument will be 
the correct reproduction of the $\overline{\mbox{MS}}$ coefficients. 
Importantly though we need only consider graphs which are built out of the {\em
single} interaction of (7). 

We now turn to the details of the calculation. Ordinarily one computes $\omega$ 
$=$ $-$ $\half\beta^\prime(g_c)$ by considering corrections to (6) in the Dyson
equations, \cite{7,13}. Equivalently one can identify the composite operator in
the lagrangian whose coupling relates to the ordinary coupling constant, 
\cite{9}.  Then the anomalous dimension of that operator is related via a 
scaling law deduced from the lagrangian to $\beta^\prime(g_c)$. In QCD the 
appropriate operator is $(G^a_{\mu\nu})^2$ as we use a formulation, (1), where 
the coupling is defined in such a way that the $3$-point interaction is 
$\bar{\psi}\gamma^\mu T^a\psi A^a_\mu$. Thus the canonical dimensions for the 
fields essentially satisfy the condition for conformal integration or 
uniqueness, \cite{19}. From the second term of (1), therefore, we have the 
scaling relation  
\begin{equation} 
\omega ~=~ \eta ~+~ \chi ~+~ \chi_G 
\end{equation} 
The quantity $\chi_G$ is the critical exponent corresponding to the 
renormalization of the pure (composite) operator $(G^a_{\mu\nu})^2$, whilst the
gluon dimension arises because of the wave function renormalization of the 
constituent fields of the composite. Thus computing $\chi_G$ gives $\omega$ 
from (8). 

To evaluate $\chi_G$ one substitutes the critical propagators (6) into the 
relevant $O(1/\Nf)$ set of Feynman diagrams and determines the residue of the
simple pole in $\Delta$, \cite{8}. This regularizing parameter is introduced
by the shift $\beta$ $\rightarrow$ $\beta$ $-$ $\Delta$. The contributing
graphs are illustrated in fig. 1 and each is computed in the Landau gauge to
avoid mixing with $(\partial^\mu A^a_\mu)^2$, \cite{9}. The first two 
graphs occur in QED and have been computed in \cite{13}. Here we note their 
respective colour group factors are $C_2(R)$ and $[C_2(R)$ $-$ $C_2(G)/2]$. The 
third graph arises from the cubic term of the operator and it and the final 
graph are purely non-abelian, each having group factors $C_2(G)$. The 
computation was carried out by the application of standard techniques for 
massless integrals including integration by parts and conformal methods. 
Nevertheless several difficult subintegrals lurk within the final graph which 
were tedious to determine. To verify that we had obtained the correct values 
for them we calculated several graphs of \cite{9} which contained the same 
subintegrals and checked that the total expression we computed agreed with the 
values given in \cite{9}. Useful in this and other respects were the packages 
{\sc Reduce} \cite{20} and {\sc Form} \cite{21}. We note that the values are 
respectively,  
\begin{eqnarray} 
&& - \, \frac{\mu(\mu-1)(2\mu-1)\eta^{\mbox{o}}_1}{(\mu+1)} ~~,~~ 
\frac{(4\mu^2+\mu-9)\eta^{\mbox{o}}_1}{(\mu+1)} ~~,~~ 
- \, \frac{(4\mu^3-2\mu^2-4\mu+1)\eta^{\mbox{o}}_1}{2(\mu+1)(2\mu-1)(\mu-2)} 
\nonumber \\ 
&& \quad \quad \quad \quad \quad \quad 
\frac{(4\mu^6-6\mu^5+18\mu^4-67\mu^3+85\mu^2-38\mu+6)\eta^{\mbox{o}}_1}
{4(2\mu-1)(\mu+1)(\mu-1)(\mu-2)} 
\end{eqnarray}  
Further at leading order there are no ghost contributions. One can see this by
attempting to include them in the formalism we will describe later and then 
observing that the first appearance of any contribution is at $O(1/\Nf^2)$. 
This feature was also observed in \cite{9} where the non-abelian generalization
of the $CP(N)$ $\sigma$ model was studied. Indeed we make use of some of the 
observations of \cite{9} here. 

The final result is 
\begin{equation} 
\omega ~=~ (\mu - 2) ~-~ \left[ (2\mu-3)(\mu-3)C_2(R) 
- \frac{(4\mu^4 - 18\mu^3 + 44\mu^2 - 45\mu + 14)C_2(G)}{4(2\mu-1)(\mu-1)} 
\right] \frac{\eta^{\mbox{o}}_1}{T(R)N_{\! f}}  
\end{equation} 
A final check on (10) is that it correctly reproduces the $O(1/\Nf)$ terms of 
the three loop $\beta$-function of (2), [1-4]. This amounts to the terms with 
$C_2(G)$ as those with $C_2(R)$ have been verified for QED in \cite{12,2}. In 
three dimensions 
\begin{equation} 
\omega ~=~ - \, \frac{1}{2} ~-~ \frac{10C_2(G)}{3\pi^2T(R)N_{\! f}} 
\end{equation} 
From (10) we can now deduce higher order coefficients which will appear in the 
$\overline{\mbox{MS}}$ $\beta$-function, by carrying out the 
$\epsilon$-expansion of (10) and using (3). Thus defining the leading order 
large $\Nf$ coefficients by $a_n$,  
\begin{equation} 
\beta(g) ~=~ \beta_0g^2 + \sum_{n=1}^\infty a_{n+1} [T(R)\Nf]^n g^{n+2} 
\end{equation}  
with $\beta_0$ $=$ $[2T(R)\Nf/3$ $-$ $11C_2(G)/6]$, then 
\begin{eqnarray} 
a_4 &=& - ~ \frac{[154C_2(R) + 53C_2(G)]}{3888} \nonumber \\ 
a_5 &=& \frac{[(288\zeta(3) + 214)C_2(R) + (480\zeta(3) - 229)C_2(G)]}{31104}
\nonumber \\ 
a_6 &=& \frac{1}{233280}[(864\zeta(4) - 1056\zeta(3) + 502)C_2(R) 
         + (1440\zeta(4) - 1264\zeta(3) - 453)C_2(G)] \nonumber \\ 
a_7 &=& \frac{1}{1679616}[(3456\zeta(5) - 3168\zeta(4) - 2464\zeta(3) 
         + 1206)C_2(R) \nonumber \\
    && ~~~~~~~~~~~~~ + \, (5760\zeta(5) - 3792\zeta(4) - 848\zeta(3) 
                   - 885)C_2(G)] 
\end{eqnarray} 

Having obtained the set $\{a_n\}$ for non-abelian theories we can examine the
purely four dimensional $\beta$-function and search for fixed points other than 
the well known infrared stable point of Banks and Zaks, $g^{BZ}_c$, \cite{14}. 
Its existence is important for recent developments in supersymmetric theories 
in relation to electric-magnetic duality, \cite{23,24}. Those infrared fixed 
points are determined by using exact non-perturbative arguments in the limit 
$N_c$, $\Nf$ $\rightarrow$ $\infty$ with $\Nf/3N_c$ held fixed, \cite{24}. (The
one loop coefficient of the $\beta$-function for that model is $(\Nf$ $-$ 
$3N_c)$ for $SU(N_c)$.) Further in the context of $1/\Nf$ expansions a 
$(16\half$ $-$ $\Nf)$ expansion from $g^{BZ}_c$ has been used to obtain an 
estimate for $\alpha_S$ for low $\Nf$, \cite{25}. Before studying the effect 
$O(1/\Nf)$ corrections have on $\beta(g)$, we first recall properties of 
$g^{BZ}_c$. In \cite{14} it was observed that for a range of $\Nf$ the two 
terms of the two loop $\beta$-function have a different sign which therefore
gives rise to a non-zero critical coupling, $g^{BZ}_c$. For $SU_c(3)$ this 
range is $8$ $<$ $\Nf$ $<$ $17$, \cite{14}, and we have recorded the explicit 
values of $g^{BZ}_c$ in this case in our notation in table 1. Subsequently one 
can also study the effect that the inclusion the three loop term of the 
$\beta$-function has on the location of $g^{BZ}_c$. We have analysed (2) 
numerically and determined the three loop values of $g^{BZ}_c$ which are given 
in the second column of table 1. Several features are apparent. First, the 
range of $\Nf$ for the existence of such an infrared fixed point is extended to
$5$ $<$ $\Nf$ $<$ $17$. Second the effect the three loop correction has is to 
move the location of $g^{BZ}_c$ towards the origin. In other words to a region 
where perturbation theory would be valid. These observations, however, ought to
be qualified. It is not clear whether this picture is meaningful because as 
$\Nf$ decreases $g^{BZ}_c$ clearly increases away from the region where 
perturbation theory is useful. In other words for low values of $\Nf$ we can 
not make a reliable statement on even, say, the {\em three} loop range of $\Nf$
for which $g^{BZ}_c$ occurs. One indication of where the perturbative picture 
may not be valid can be deduced from the values of the critical exponent 
$\beta^\prime(g^{BZ}_c)$ which is a physically meaningful and calculable 
quantity. In table 2 we have given the corresponding values for 
$\beta^\prime(g^{BZ}_c)$ as deduced from the two and three loop values of 
$g^{BZ}_c$ respectively. Clearly the three loop corrections do not affect the 
two loop values appreciably for $\Nf$ $=$ $14$, $15$ and $16$ suggesting that 
higher loop corrections are small. For lower values the divergence is evident 
indicating that the four and higher loop contributions would be needed to make
an accurate estimate of the exponent. In light of the critical coupling being 
smaller for a larger range of $\Nf$ it would better of either, though, to take 
the three loop values of $\beta^\prime(g^{BZ}_c)$ as being the more reliable.  

Now we consider the effect that the $O(1/\Nf)$ corrections of (10) have. We have
studied the case $N_c$ $=$ $3$ in various ways. First, as in \cite{12} we 
examined the $\beta$-function given by just taking all the leading order 
coefficients $a_n$ which was improved by non-abelianization, \cite{26}. This
entails replacing $\Nf$ by the one loop $\beta$-function coefficient through 
the shift $\Nf$ $\rightarrow$ $(\Nf$ $-$ $11C_2(G)/[4T(R)])$. It turns out that
in searching for zeroes of the four dimensional $\beta$-function that the 
contributions from these $O(1/\beta_0)$ coefficients on their own are not 
sufficient for even obtaining a fixed point $g^{BZ}_c$. This is the same as was
found in the QED case, \cite{12}, in the range of couplings where the series 
was convergent. Instead, to improve this situation we took the two and three
loop $\beta$-functions of (2) and then included all subsequent information 
included in (10). The point of view being that one can at least study the 
effect the $O(1/\Nf)$ corrections have on the fixed point $g^{BZ}_c$ which is 
known to exist. It turns out that in this approach we did not observe any 
non-trivial fixed points other than $g^{BZ}_c$ in $g$ $>$ $0$ which was 
independent of the number of terms included. From a practical point of view in 
our analysis we truncated the series for $\beta(g)$ at around $14$ terms. The 
effect of including more terms is negligible on the results we give in both 
tables until $\Nf$ $\leqq$ $8$ when perturbation theory can not be regarded as 
reliable anyway. The remaining columns of our tables are the results of this 
analysis. Clearly the effect the $O(1/\beta_0)$ corrections has is not to move 
$g^{BZ}_c$ significantly from the three loop value for a large range of $\Nf$. 
Also for $\Nf$ $=$ $14$, $15$ and $16$ the values of $\beta^\prime(g^{BZ}_c)$ 
are not that different from the perturbative estimates of 
$\beta^\prime(g^{BZ}_c)$. 

In conclusion we have produced the leading order corrections to the QCD 
$\beta$-function in a $1/\Nf$ expansion which extends the calculation of 
\cite{12}. Consequently we examined the effect they had on the known infrared 
fixed point in four dimensional $\beta$-function. It transpires that 
perturbation theory is valid for analysing the fixed point when the value of 
$\Nf$ is near the upper bound for the existence of $g^{BZ}_c$ and estimates 
for a critical exponent were obtained then. It would be interesting, though, to
compare the values of these exponents with results from other techniques such 
as the lattice, which would be expected to reliably cover the lower part of the
range.  In this case a resummation would be necessary to try and improve the 
lack of convergence which is apparent when the three loop values are compared 
to the two loop ones. Further, we believe it would be useful to repeat our 
analysis for the supersymmetric extension of QCD in large $\Nf$ in relation to 
\cite{23,24}. Once the analogous expression to (10) is available it would be
possible to study the effect the $O(1/\Nf)$ corrections have on the infrared
fixed point when $N_c$ is large as well as for orthogonal and symplectic 
gauge groups. As a first step one would need to determine which field theory 
supersymmetric QCD is equivalent to at the $d$-dimensional fixed point and 
verify, for example, that the correct triple and quartic interactions are 
obtained in the large $\Nf$ limit similar to \cite{8}. It would be hoped that 
there is a small set of interactions, as in the non-abelian Thirring model, to 
reduce the amount of calculation that would occur. 

\noindent 
{\bf Acknowledgements.} This work was carried out with the support of PPARC 
through an Advanced Fellowship. The author thanks Drs D.J. Broadhurst, D.R.T.
Jones and H. Osborn for useful conversations and Dr T.J. Morris for drawing his
attention to \cite{18}. The figures were designed using the package {\sc 
FeynDiagram} version 1.21.

\newpage 
{\begin{table} 
\begin{center} 
\begin{tabular}{c||c|c|c|c} 
$N_{\!f}$ & Two loop & Three loop & Two loop & Three loop \\ 
&&&$+$ $O(1/\beta_0)$ &$+$ $O(1/\beta_0)$ \\ 
\hline 
 6 & - & 4.050746 & (0.803122) & 0.517376 \\ 
 7 & - & 0.782073 & (0.829257) & 0.463261 \\ 
 8 & - & 0.466021 & (0.858331) & 0.388537 \\ 
 9 & 1.666667 & 0.327211 & 0.878128 & 0.308364 \\ 
10 & 0.702703 & 0.243297 & 0.858362 & 0.238810 \\  
11 & 0.392857 & 0.184156 & 0.666176 & 0.183119 \\ 
12 & 0.240000 & 0.138426 & 0.283504 & 0.138208 \\ 
13 & 0.148936 & 0.100763 & 0.155791 & 0.100726 \\ 
14 & 0.088496 & 0.068282 & 0.089480 & 0.068278 \\ 
15 & 0.045455 & 0.039271 & 0.045534 & 0.039270 \\ 
16 & 0.013245 & 0.012647 & 0.013246 & 0.012647 \\ 
\end{tabular} 
\end{center} 
\begin{center} 
{Table 1. Location of infrared fixed point $g^{BZ}_c$ for $SU_c(3)$.} 
\end{center} 
\end{table}}  

{\begin{table} 
\begin{center} 
\begin{tabular}{c||c|c|c|c} 
$N_{\!f}$ & Two loop & Three loop & Two loop & Three loop \\ 
&&&$+$ $O(1/\beta_0)$ &$+$ $O(1/\beta_0)$ \\ 
\hline 
 6 & - & 81.682972 & (15.663662) & 8.095948 \\ 
 7 & - & 5.972522 & (17.792600) & 4.922952 \\ 
 8 & - & 2.658882 & (16.790094) & 2.717457 \\ 
 9 & 4.166667 & 1.475455 & 13.456718 & 1.510817 \\ 
10 & 1.522523 & 0.871775 & 7.901656 & 0.879995 \\ 
11 & 0.720238 & 0.516977 & 1.617025 & 0.518561 \\ 
12 & 0.360000 & 0.295517 & 0.360750 & 0.295784 \\ 
13 & 0.173759 & 0.155581 & 0.173789 & 0.155616 \\ 
14 & 0.073746 & 0.069899 & 0.073751 & 0.069903 \\ 
15 & 0.022727 & 0.022307 & 0.022727 & 0.022306 \\ 
16 & 0.002208 & 0.002203 & 0.002207 & 0.002203 \\ 
\end{tabular} 
\end{center} 
\end{table} 
\begin{center} 
{Table 2. Values of $\beta^\prime(g^{BZ}_c)$ for $SU_c(3)$.} 
\end{center}}  

\newpage 
\epsfxsize=18cm 
\epsfbox{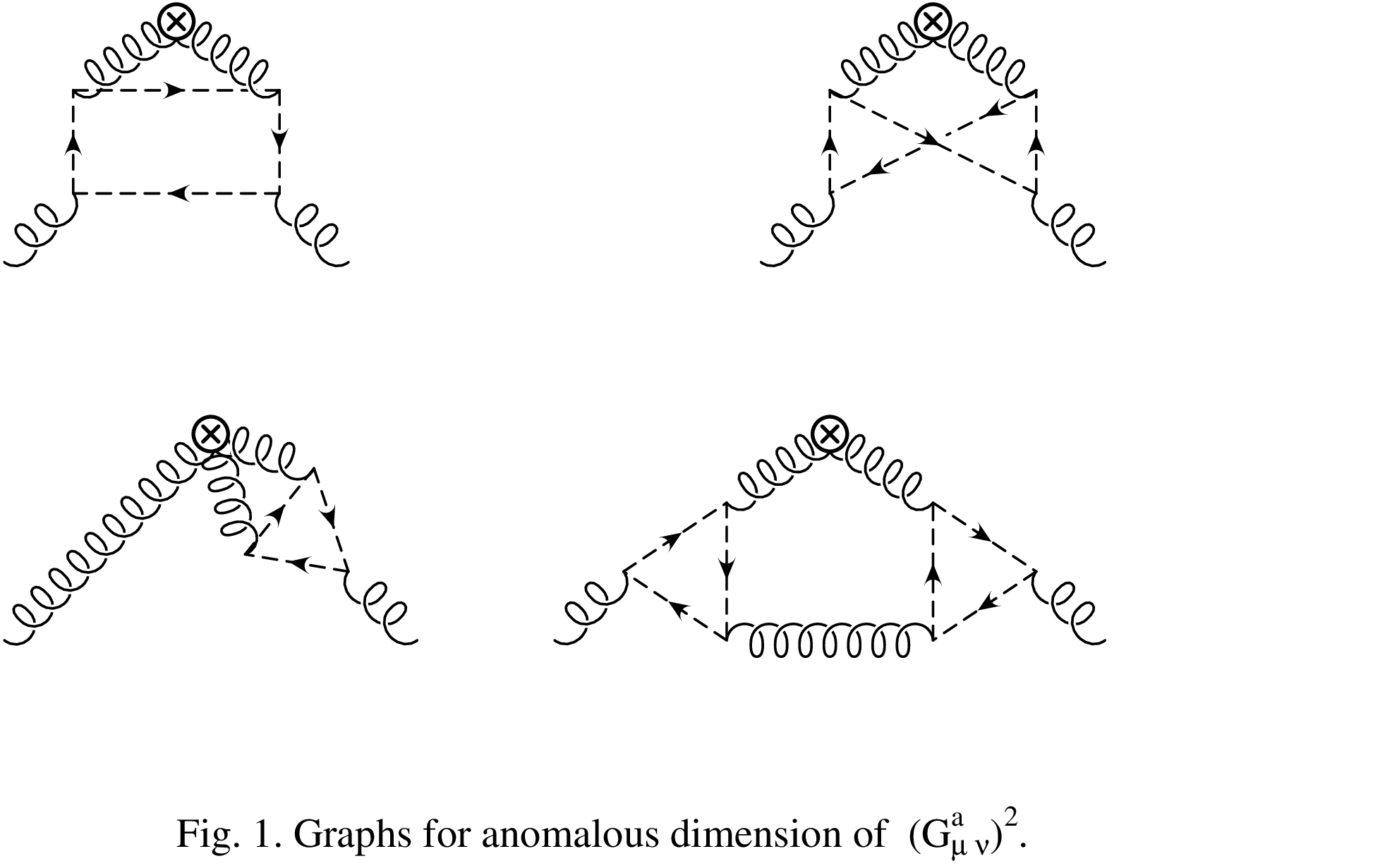} 


\newpage 
\begin{thebibliography}{99}
\bibitem{1} D.J. Gross \& F.J. Wilczek, Phys. Rev. Lett. {\bf 30} (1973),
1343; H.D. Politzer, Phys. Rev. Lett. {\bf 30} (1973), 1346.
\bibitem{2} W.E. Caswell, Phys. Rev. Lett. {\bf 33} (1974), 244; D.R.T.
Jones, Nucl. Phys. {\bf B75} (1974), 531. 
\bibitem{3} O.V. Tarasov, A.A. Vladimirov \& A.Yu. Zharkov, Phys. Lett. 
{\bf 93B} (1980), 429.
\bibitem{4} S.A. Larin \& J.A.M. Vermaseren, Phys. Lett. {\bf B303} (1993),
334.
\bibitem{5} D.V. Nanopoulos \& D.A. Ross, Nucl. Phys. {\bf B157} (1979), 273;  
R. Tarrach, Nucl. Phys. {\bf B183} (1981), 384. 
\bibitem{6} O.V. Tarasov, JINR preprint P2-82-900 (in Russian).
\bibitem{7} A.N. Vasil'ev, Yu.M. Pis'mak \& J.R. Honkonen, Theor. Math. Phys.
{\bf 46} (1981), 157; {\em ibid.} {\bf 47} (1981), 291. 
\bibitem{8} A.N. Vasil'ev \& M.Yu. Nalimov, Theor. Math. Phys. {\bf 55} (1982),
423; {\em ibid.} {\bf 56} (1982), 643.
\bibitem{9} A.N. Vasil'ev, M.Yu. Nalimov \& J.R. Honkonen, Theor. Math. Phys. 
{\bf 58} (1984), 111. 
\bibitem{10} J.A. Gracey, Phys. Lett. {\bf B318} (1993), 177. 
\bibitem{11} E.S. Egorian \& O.V. Tarasov, Theor. Math. Phys. {\bf 41} (1979),
26.
\bibitem{12} A. Palanques-Mestre \& P. Pascual, Commun. Math. Phys. {\bf 95}
(1984), 277. 
\bibitem{13} J.A. Gracey, Int. J. Mod. Phys. {\bf A8} (1993), 2465.
\bibitem{14} T. Banks \& A. Zaks, Nucl. Phys. {\bf B196} (1982), 189. 
\bibitem{15} J. Zinn-Justin, Nucl. Phys. {\bf B367} (1991), 105. 
\bibitem{16} S.J. Hands, Phys. Rev. {\bf D51} (1995), 5816. 
\bibitem{17} K.-I. Kondo, Nucl. Phys. {\bf B450} (1995), 251. 
\bibitem{18} A. Hasenfratz \& P. Hasenfratz, Phys. Lett. {\bf B297} (1992), 
166. 
\bibitem{19} M. d'Eramo, L. Peliti \& G. Parisi, Lett. Nuovo Cim. {\bf 2}
(1971), 878.
\bibitem{20} A.C. Hearn, ``{\sc REDUCE} Users Manual'' version 3.4, Rand 
publication CP78, (1991). 
\bibitem{21} J.A.M. Vermaseren, ``{\sc FORM}'' version 1.1, CAN publication,  
(1992). 
\bibitem{22} S.G. Gorishny, A.L. Kataev, S.A. Larin \& L.R. Surguladze, Phys.
Lett. {\bf B256} (1991), 81. 
\bibitem{23} N. Seiberg \& E. Witten, Nucl. Phys. {\bf B426} (1994), 19. 
\bibitem{24} N. Seiberg, Nucl. Phys. {\bf B435} (1995), 129. 
\bibitem{25} P.M. Stevenson, Phys. Lett. {\bf B331} (1994), 187. 
\bibitem{26} D.J. Broadhurst \& A.G. Grozin, Phys. Rev. {\bf D52} (1995), 4082. 
\end{thebibliography}
\end{document}